\documentstyle[psfig,aps,multicol,eqsecnum]{revtex}
\newcommand{\be}{\begin{eqnarray}}
\newcommand{\ee}{\end{eqnarray}}
\newcommand{\la}{\langle}
\newcommand{\ra}{\rangle}
\newcommand{\tra}{{\rm Tr}} 

\newcommand{\sq}{\sqrt q}
\newcommand{\sqq}{\sqrt{1-q}}
\begin{document}
\title{On the von Neumann capacity of noisy quantum channels} 
\author{C. Adami$^{1,2,3}$ and N. J. Cerf$^{1,3}$} 
\address{$^1$W. K. Kellogg Radiation Laboratory
and $^2$Computation and Neural Systems\\ California Institute of
Technology, Pasadena, California 91125\\$^3$Institute for Theoretical
Physics, University of California, Santa Barbara, California 93106}

\date{Received 15 October 1996}

\draft
\maketitle
\begin{abstract}
  We discuss the capacity of quantum channels for information
  transmission and storage.  Quantum channels have dual uses: they can
  be used to transmit {\em known} quantum states which code for
  classical information, and they can be used in a purely quantum
  manner, for transmitting or storing quantum entanglement.  
  We propose here a definition of the {\em von Neumann} capacity of
  quantum channels, which is a quantum mechanical {\em extension} of
  the Shannon capacity and reverts to it in the classical limit. As
  such, the von Neumann capacity assumes the role of a classical or
  quantum capacity depending on the usage of the channel. In analogy
  to the classical construction, this capacity is defined as the
  maximum {\em von Neumann mutual entropy} processed by the channel, a
  measure which reduces to the capacity for classical information
  transmission through quantum channels (the ``Kholevo capacity'')
  when {\em known} quantum states are sent.  The quantum mutual
  entropy fulfills all basic requirements for a measure of
  information, and observes quantum data-processing inequalities. 
  We also derive a quantum Fano inequality relating the {\em quantum
    loss} of the channel to the fidelity of the quantum code.  The
  quantities introduced are calculated explicitly for the quantum
  ``depolarizing'' channel.  The von Neumann capacity is interpreted
  within the context of superdense coding, and an ``extended'' Hamming
  bound is derived that is consistent with that capacity.

\end{abstract}
\pacs{PACS numbers: 03.65.Bz,89.70.+c
      \hfill KRL preprint MAP-206}
\begin{multicols}{2}[]
\narrowtext
\section{Introduction}
The problem of transmission and storage of quantum states has received
a considerable amount of attention recently, owing to the flurry of
activity in the field of quantum computation~\cite{bib_reviews}
sparked by Shor's discovery of a quantum algorithm for
factoring~\cite{bib_shor}. In anticipation of physical realizations of
such computers (which still face major conceptual challenges), it is
necessary to extend to the quantum regime the main results of
Shannon's information theory~\cite{bib_shannon}, which provides limits
on how well information can be compressed, transmitted, and preserved.
In this spirit, the quantum analogue of the noiseless coding theorem
was obtained recently by Schumacher~\cite{bib_schum}.  However, noisy
quantum channels are less well understood, mainly because quantum
noise is of a very different nature than classical noise, and the
notion of ``quantum information'' is still under discussion. Yet,
important results have been obtained concerning the correction of
errors induced by the decoherence of quantum bits via suitable quantum
codes. These error-correcting
codes~\cite{bib_shor1,bib_calder,bib_steane,bib_laflamme,bib_ekert,bib_bdsw,bib_knill,bib_calder1}
work on the principle that quantum information can be encoded in
blocks of qubits (codewords) such that the decoherence of any qubit
can be corrected by an appropriate code, much like the classical
error-correcting codes. Therefore, it is expected that a
generalization of Shannon's fundamental theorem to the quantum regime
should exist, and efforts towards such a proof have appeared
recently~\cite{bib_schum1,bib_schum2,bib_lloyd}. The capacity for the
transmission of {\em classical} information through quantum channels
was recently obtained by Hausladen {\it et al.}~\cite{bib_hausladen}
for the transmission of pure states, and by
Kholevo~\cite{bib_kholevo97} for the general case of mixed states.

When discussing quantum channels, it is important to keep in mind that
they can be used in two very different modes.  On the one hand, one
may be interested in the capacity of a channel to transmit or else
store, an {\em unknown} quantum state in the presence of quantum
noise. This mode is unlike any use of a channel we are accustomed to
in classical theory, as strictly speaking classical information is not
transmitted in such a use (no measurement is involved).  Rather, such
a capacity appears to be a measure of how much {\em entanglement} can
be transmitted (or maintained) in the presence of noise induced by the
interaction of the quantum state with a ``depolarizing'' environment.
On the other hand, a quantum channel can be used for the transmission
of {\em known} quantum states (classical information), and the
resulting capacity (i.e., the classical information transmission
capacity of the quantum channel) represents the usual bound on the
rate of arbitrarily accurate information transmission.  In this paper,
we propose a definition for the {\em von Neumann} capacity of a
quantum channel, which encompasses the capacity for procesing quantum
as well as classical information.  This definition is based on a
quantum mechanical extension of the usual Shannon mutual entropy to a
von Neumann mutual entropy, which measures quantum as well as
classical correlations.  Still, a natural separation of the von
Neumann capacity into classical and purely quantum pieces does not appear to
be straightforward.  This reflects the difficulty in separating
classical correlation from quantum entanglement (the ``quantum
separability'' problem, see, e.g., \cite{bib_horo} and references
therein).  It may be that there is no unambiguous way to separate
classical from purely quantum capacity for all channels and all noise
models.  The von Neumann capacity we propose, as it does not involve
such a separation, conforms to a number of ``axioms'' for such a
measure among which are positivity, subadditivity, concavity
(convexity) in the input (output), as well as the data processing
inequalities.  We also show that the von Neumann capacity naturally
reverts to the capacity for classical information transmission through
noisy quantum channels of Kholevo~\cite{bib_kholevo97} (the Kholevo
capacity) if the unknown states are measured just before transmission,
or, equivalently, if the quantum states are {\em prepared}. In such a
use, thus, the ``purely quantum piece'' of the von Neumann capacity
vanishes. We stop short of proving that the von Neumann capacity can
be achieved by quantum coding, i.e., we do not prove the quantum
equivalent of Shannon's noisy coding theorem for the total capacity.
We do, however, provide an example where the von Neumann capacity
appears achievable: the case of noisy superdense coding.

In the next section we recapitulate the treatment of the {\em
  classical} communication channel in a somewhat novel manner, by
insisting on the deterministic nature of classical physics with
respect to the treatment of information.  This treatment paves the way
for the formal discussion of quantum channels along the lines of
Schumacher~\cite{bib_schum1} in Section III, which results in a
proposal for the definition of a von Neumann capacity for transmission
of entanglement/correlation that parallels the classical construction.
We also prove a number of properties of such a measure, such as
subadditivity, concavity/convexity, forward/backward quantum
data-processing inequalities, and derive a quantum Fano inequality
relating the loss of entanglement in the channel to the fidelity of
the code used to protect the quantum state. This proof uses an
inequality of the Fano-type obtained recently by
Schumacher~\cite{bib_schum1}.  In Section IV we demonstrate that the
von Neumann capacity reduces to the recently obtained Kholevo
capacity~\cite{bib_kholevo97} if the quantum states are {\em known},
i.e., measured and ``kept in memory'', before sending them on.  In
Section V then we apply these results directly to a specific example,
the quantum depolarizing channel~\cite{bib_channel}. This generic
example allows a direct calculation of all quantities involved.
Specifically, we calculate the entanglement/correlation processed by
the channel as a function of the entropy of the input and the
probability of error of the channel.  We also show that this capacity
reverts to the well-known capacity for classical information
transmission in a depolarizing channel if {\em known} quantum states
are transmitted through the channel.  In Section VI finally, we
interpret the von Neumann capacity in the context of superdense coding
and derive a quantum Hamming bound consistent with it.

\section{Classical channels}
The information theory of classical channels is well known since
Shannon's seminal work on the matter~\cite{bib_shannon}. In this
section, rather than deriving any new results, we expose
the information theory of classical channels in the light of the {\em
  physics} of information, in preparation of the quantum treatment of
channels that follows.  Physicists are used to classical laws of
physics that are {\em deterministic}, and therefore do not consider
noise to be an intrinsic property of channels. In other words,
randomness, or a stochastic component, does not exist {\em per se},
but is a result of incomplete measurement. Thus, for a physicist there
are no noisy channels, only incompletely monitored ones. As an
example, consider an information transmission channel where the
sender's information is the face of a coin before it is flipped, and
the receiver's symbol is the face of the coin after it is flipped.
Information theory would classify this as a useless channel, but for a
physicist it is just a question of knowing the initial conditions of
the channel {\em and the environment} well enough. From this, he can
calculate the trajectory of the coin, and by examining the face at the
received side infer the information sent by the sender. Classical
physics, therefore, demands that all {\em conditional} probability
distributions can be made to be {\em peaked}, if the environment,
enlarged enough to cover all interacting systems, is monitored.  In
other words, $p_{i|j}=1$ or $0$ for all $i$, $j$: if the
outcome $j$ is known, $i$ can be inferred with certainty. 
As a consequence, {\em  all} conditional entropies can be made to 
vanish for a closed system.

\begin{figure}
\caption{(a) Entropy Venn diagram for the classical channel
$XY$, and its ``physical'' extension including the environment.}
\vskip 0.25cm
\centerline{\psfig{figure=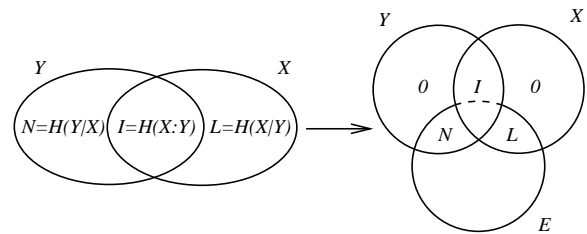,width=3.0in,angle=-90}}
\label{fig_class}
\vskip -0.25cm
\end{figure}

According to this principle, let us then construct the
classical channel. Along with the ensemble of source symbols $X$
(symbols $x_1,\cdots,x_N$ appearing with probabilities
$p_1,\cdots,p_N$), imagine an ensemble of received symbols $Y$. The
usual noisy channel is represented by the diagram on the left in
Fig.~\ref{fig_class}: the conditional entropy $H(X|Y)$ represents the
loss $L$ in the channel, i.e., the uncertainty of inferrring $X$ from
$Y$, whereas $H(Y|X)$ stands for noise $N$ in the output,
which is unrelated to the error-rate of the channel. 
A channel for which $L=0$ is called a ``lossless'' channel (no
transmission errors occur), whereas $N=0$ characterizes a
``deterministic'' channel (the input unambiguously determines the
output). On the right-hand side in Fig.~\ref{fig_class}, we have
extended the channel to include the environment. All conditional
entropies are zero, and the noise and loss are simply due to
correlations of the source or received ensembles with an
environment, i.e., $L=H(X{\rm:}E|Y)$ and $N=H(Y{\rm:}E|X)$.  
The capacity of the classical channel is obtained by
maximizing the mutual entropy between source and received symbols [the
information $I=H(X{\rm:}Y)$ processed by the channel] over all input 
distributions: 
\be
C = \max_{p(x)}\; I\;. 
 \ee 
If the output of the channel $Y$ is
subjected to {\em another} channel (resulting in the output $Z$, say),
it can be shown that the information processed by the combined channel,
$H(X{\rm:}Z)$, cannot possibly be larger than the information processed 
in the {\em first} leg, $H(X{\rm:}Y)$. In other words, any subsequent 
processing of the output 
cannot possibly increase the transmitted information. This is
expressed in the 
data-processing inequality (see, e.g.,~\cite{bib_ash}):
\be
H(X{\rm:}Z)\leq H(X{\rm:}Y)\leq H(X) \label{dataproc}\;.
\ee 
On the same token, a ``reverse'' data-processing inequality can be
proven, which implies that the information processed in the {\em second}
leg of the channel, $H(Y{\rm:}Z)$, must exceed the information
processed by the total channel, $H(X{\rm:}Z)$:
\be
H(X{\rm:}Z)\leq H(Y{\rm:}Z)\leq H(Z) \label{dataproc1}\;.
\ee
This inequality reflects microscopic time-reversal invariance: any
channel used in a forward manner can be used in a backward manner.

As far as
coding is concerned, the troublesome quantity is the loss $L$, while 
the noise $N$ is unimportant.
Indeed, for a message of length $n$, the typical number of input
sequences for every output sequence is $2^{nL}$, making decoding
impossible.  The principle of error-correction is to embed the messages
into {\em codewords}, that are chosen in such a way that the
conditional entropy of the ensemble of codewords {\em vanishes}, i.e.,
on the level of message transmission the channel is lossless. Not
surprisingly, there is then a relationship between the channel loss
$L$ and the probability of error $p_c$ of a {\em code} $c$ 
that is composed of $s$ codewords: 
\be 
L\leq H_2[p_c]+p_c\log(s-1)\;,  \label{fanocl}
\ee
where $H_2[p]$ is the dyadic Shannon entropy 
\be
H_2[p] = H_2[1-p] = -p\log p\,-\,(1-p)\log(1-p)\;.
\ee
Eq.~(\ref{fanocl}) is the Fano inequality (see, e.g.,~\cite{bib_ash}), 
which implies, for example, that the loss vanishes 
if the error of the code vanishes. Note that the noise of the 
channel itself in general is not zero in this situation. Let us now turn to 
quantum channels.

\section{Quantum channels}
\subsection{Information theory of entanglement}
Quantum channels have properties fundamentally different from the
classical channel just described owing to the superposition principle
of quantum mechanics and the non-cloning theorem that
ensues~\cite{bib_nocloning}. First and foremost, the ``input'' quantum
state, after interaction with an environment, is ``lost'', having
become the output state. Any attempt at copying the quantum state
before decoherence will result in a classical channel, as we
will see later. Thus, a joint probability for input and output symbols
does not exist for quantum channels. However, this is not essential as
the quantity of interest in quantum communication is {\em not} the
state of an isolated quantum system (a ``product state''), but the
degree of entanglement between one quantum system and another,
parameterized by their mutual entropy as shown below. A
single non-entangled quantum system (such as an isolated spin-1/2
state) carries no entropy and is of no interest for quantum
communication as it can be arbitrarily recreated at any time.  Entangled
composite systems (such as Bell states) on the other hand are
interesting because the entanglement can be used for communication.
Let us very briefly recapitulate the quantum
information theory of
entanglement~\cite{bib_neginfo,bib_entang,bib_meas,bib_reality}.

\begin{figure}
\caption{(a) Entropy Venn diagram for a bipartite entangled quantum
system $AB$, depicting $S(AB)$ (total area), marginal entropies
[$S(A)$ viz. $S(B)$], conditional [$S(A|B)$ viz. $S(B|A)$] and mutual
[$S(A{\rm:}B)$] entropies. (b) Entropy diagram for a fully entangled
Bell-state.}
\vskip 0.3cm
\centerline{\psfig{figure=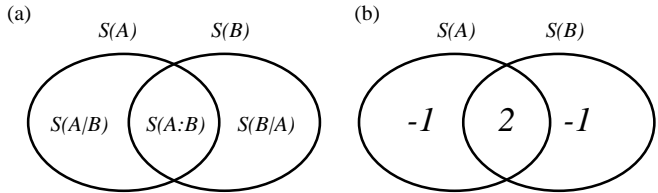,width=3.4in,angle=0}}
\label{fig_venn}
\vskip 0.25cm
\end{figure}

For a composite quantum system $AB$, we can write relations between
von Neumann entropies that precisely parallel those written by Shannon for 
classical entropies. Specifically, we can define the conditional entropy 
of $A$ (conditional on the knowledge of $B$)
\be
S(A|B) = S(AB)-S(B)
\ee
via a suitable definition of a ``conditional'' density matrix
$\rho_{A|B}$.  The latter matrix can have eigenvalues larger than
unity, revealing its non-classical nature and allowing conditional quantum
entropies to be {\em negative}~\cite{bib_neginfo}. Similarly, we can define 
a ``mutual'' density matrix $\rho_{A{\rm:}B}$ giving rise to a mutual von 
Neumann entropy
\be
S(A{\rm:}B) = S(A) + S(B) - S(AB)
\ee 
which exceeds the usual bound obtained for mutual Shannon entropies
by a factor of two:
\be 
S(A{\rm:}B) \le 2\,{\rm min}[S(A),S(B)]\;.
\ee
The latter equation demonstrates that quantum systems can be more
strongly correlated than classical ones: they can be {\em
supercorrelated}.  These relations can be conveniently summarized by
entropy Venn diagrams (Fig.~\ref{fig_venn}a) as is usual in classical 
information
theory. The extension to the quantum regime implies that negative
numbers can appear which are classically forbidden\footnote{
In classical entropy Venn diagrams, negative numbers can only 
appear in the mutual entropy of three or more systems.}.
As an example, we show in Fig.~\ref{fig_venn}b the quantum entropies of Bell
states (which are fully entangled states of two qubits). These notions
can be extended to multipartite systems, and will be used throughout
the paper.  

The degree of entanglement of a bipartite pure quantum state is
customarily indicated by the marginal entropy of one of its parts,
i.e., the von Neumann entropy of the density matrix obtained by
tracing the joint density matrix over the degrees of freedom of the
other part (the entropy of entanglement, see~\cite{bib_bdsw}).  
However, since the parts of an entangled system 
do not possess a state on their own, it takes up to twice the
marginal entropy of one of the parts to specify (in bits) the state of
entanglement.  For example, it takes up to two bits to specify the
entanglement between two qubits (there are four Bell-basis
states). Thus, we propose to measure the entanglement of pure states 
by the mutual entropy between the two parts, which takes values between 0 (for
non-entangled systems) and $2S$ (for entangled systems of
marginal entropy $S$ each). In order to avoid confusion with the previously
defined entropy of entanglement, we propose to call this quantity
the {\em mutual entanglement} (or simply von Neumann mutual entropy), 
and denote it by the symbol $I_Q$:
\be
I_Q = S(A{\rm:}B)\;.
\ee
For pure entangled states, the mutual
entanglement $I_Q$ is just twice the entropy of entanglement, 
demonstrating
that either is a good measure for the {\em degree} of entanglement,
but not necessarily for the absolute amount. Estimating the
entanglement of {\em mixed} states, on the other hand, is more
complicated, and no satisfying definition is available
(see~\cite{bib_bdsw} for the
most established ones). The quantum mutual entropy for mixed states 
does {\em not} represent pure quantum entanglement, but rather 
classical {\em and} quantum entanglement that is difficult to
separate consistently. For reasons that become more clear in the 
following, we believe that the mutual
entanglement $I_Q$ between two systems is the most
straightforward generalization 
of the mutual information $I$ of classical information theory, and
will serve as the vehicle to define a quantum/classical {\em von
Neumann} capacity for quantum channels.
\subsection{Explicit model}
\begin{figure}
\caption{Quantum network representation of a noisy quantum
  channel. $R$ purifies the mixed state $Q$; the corresponding
  entanglement is indicated by a dashed line.
\label{fig_channel}}
\vskip 0.25cm
\centerline{\psfig{figure=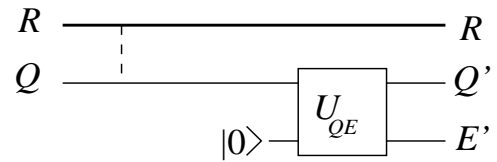,width=2.5in,angle=-90}}
\vskip -0.25cm
\end{figure}

\begin{figure}
\caption{Unitary transformation entangling the pure environment
$|E\ra$ with the pure system $|QR\ra$. The reference system $R$ is not
touched by this transformation, which implies that no entropy can be exchanged 
across the double solid lines in the diagram on the left.}
\vskip 0.25cm
\centerline{\psfig{figure=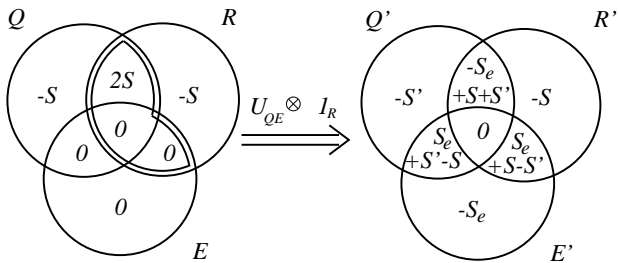,width=3.2in,angle=-90}}
\label{fig_unitary}
\vskip -0.25cm
\end{figure}

In constructing a general quantum channel formally, we follow 
Schumacher~\cite{bib_schum1}. A quantum mixed state $Q$ suffers
entanglement with an environment $E$ so as to lead to a new mixed
state $Q'$ with possibly increased or decreased entropy. In order to 
monitor the
entanglement transmission, the initial mixed state $Q$ is ``purified''
by considering its entanglement with a ``reference'' system
$R$:
\be
|RQ\ra=\sum_i\sqrt{p_i}\,|r_i,i\ra \label{eq9}
\ee
where $|r_i\ra$ are the $R$ eigenstates. 
Indeed, this can always be achieved via a Schmidt decomposition.
Then, the mixed state $Q$ is simply obtained as a partial trace of the
pure state $QR$:
\be
\rho_Q = \tra_R[\rho_{QR}]=\sum_i p_i\,|i\ra\la i|\;.\label{eq10}
\ee 
Also, the interaction with the environment
\be
QRE\stackrel {U_{QE}\otimes 1_R}\longrightarrow Q'R'E' \label{eq11}
\ee
now can be viewed as a channel to transmit the entanglement between $QR$ to the
system $Q'R'$. Here, $U_{QE}$ is the unitary operation entangling $QR$
with the environment $E$, which is initially in a pure state. This
construction is summarized in Fig.~\ref{fig_channel}. 
The
evolution of entropies in such a channel is depicted in
Fig.~\ref{fig_unitary},  where the entropy of the 
reference state [which is the same as the entropy of $Q$ {\em before}
entanglement, $S(Q)=S(R)$] is denoted by $S$, 
\be
S= -\sum_i p_i\log p_i\;, \label{eq12}
\ee
while the entropy of the quantum state $Q'$ 
after entanglement $S(Q')=S'$, and the entropy of the environment $S(E)=S_e$. 
The latter was termed ``exchange entropy'' by Schumacher~\cite{bib_schum1}. 
Note that, as for any tripartite pure state, the entropy dia\-gram of the
entangled 
state $Q'R'E'$ is uniquely fixed by three parameters, the marginal entropies
of $Q'$, $R'$, and $E'$ respectively, i.e., the numbers $S$, $S'$, and $S_e$. 
Also, in any pure entangled diagram involving three systems, the 
ternary mutual entropy [the center of the ternary diagram, 
$S(Q'{\rm:}R'{\rm:}E')$], 
is always zero~\cite{bib_entang,bib_meas,bib_reality}. 

\begin{figure}
\caption{Entropy diagram summarizing the entropy relations between the
entangled systems $Q'$, $R'$, and $E'$.  }
\vskip 0.25cm
\centerline{\psfig{figure=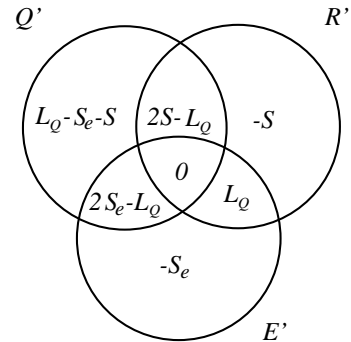,width=1.75in,angle=-90}}
\label{fig_loss}
\vskip -0.25cm
\end{figure}

To make contact with the classical channel of the previous section,
let us define the {\em quantum loss} $L_Q$\footnote{We follow here
  the nomenclature that ``quantum'' always means ``quantum including
  classical'', rather than ``purely quantum'', in the same sense as
  the von Neumann entropy is not just a purely quantum entropy.  This
  nomenclature is motivated by the difficulty to separate classical
  from quantum entanglement.}:
\be 
L_Q= S(R'{\rm:}E'|Q')=S_e+S-S'\;.
\ee
It represents the difference between the entropy acquired by the environment, 
$S_e$, and the entropy change of $Q$, ($S'-S$), and thus stands for the 
loss of entanglement in the quantum transmission. It 
plays a central role in error correction as shown below and in Section III.D. 
The entropy diagram in terms of
$S$, $S_e$, and $L_Q$ is depicted in Fig.~\ref{fig_loss}. From this diagram 
we can immediately read off inequalities relating the loss $L_Q$ and the 
entropies $S$ and $S_e$ by considering triangle
inequalities for quantum entropies~\cite{bib_araki}, namely
\be
0&\le&L_Q\le 2S \;\label{lossbound},\label{ineq1}\\
0&\le&L_Q\le 2S_e\label{ineq2}\;,
\ee
which can be combined to
\be
0\le L_Q\le 2 \min\,(S,S_e)\;.
\ee
We find therefore that the initial mutual entanglement $2S$ is split, 
through the action of the environment, into a piece shared with $Q'$ 
[i.e., $S(Q'{\rm:}R')=2S-L_Q$], 
and a piece shared with the environment (the remaining loss $L_Q$) 
according to the relation
\be
S(R'{\rm:}Q')+S(R'{\rm:}E'|Q')=S(R'{\rm:}E'Q')=S(R{\rm:}Q)\;,
\ee
or equivalently
\be
I_Q +L_Q = 2S\;.
\ee
Finally, we are ready to propose a definition for the 
von Neumann capacity. Again, in analogy 
with the classical construction, the von Neumann capacity $C_Q$ would
be the mutual entanglement processed by the channel (mutual von
Neumann entropy), maximized over 
the density matrix of the input channel, i.e.,
\be
C_Q = \max_{\rho_Q} I_Q\;,\label{quantcap}
\ee
where $I_Q=S(R'{\rm:}Q')=S(R{\rm:}Q')$ is the entanglement processed 
by the channel:
\be 
I_Q=2S-L_Q\;.  
\ee 
From the bound (\ref{lossbound}) we find that
the entanglement processed by the channel is non-negative, and bounded
from above by the initial entanglement $2S$. An interesting situation
arises when the entanglement processed by the channel saturates
this upper bound. This is the case of the {\em lossless} quantum
channel, where $L_Q=0$. 

\begin{figure}
\caption{Entanglement between $Q'$, $R'$, and $E'$ in the lossless quantum 
channel}
\vskip 0.25cm
\centerline{\psfig{figure=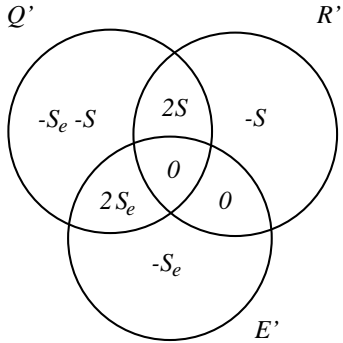,width=1.75in,angle=-90}}
\label{fig_lossless}
\vskip -0.25cm
\end{figure}

It was shown recently by Schumacher and Nielsen~\cite{bib_schum2} that
an error-correction procedure meant to restore the initial quantum
state (and thus the initial entanglement $2S$) can only be successful
when $L_Q=0$. From Fig.~\ref{fig_loss} we can see that when $L_Q=0$,
$Q'$ is entangled {\em separately} with the reference state and the
environment, leading to the diagram represented in
Fig.~\ref{fig_lossless}.  For this reason alone it is possible to
recover the initial entanglement between $Q$ and $R$ via interaction
with an ancilla $A$ (that can be viewed as a {\em second} environment
in a ``chained'' channel).  The latter effects a transfer of the
entanglement between $Q'$ and $E'$ to entanglement between $E'$ and
$A$. This operation can be viewed as an ``incomplete'' measurement of
$Q'$ by $A$ which only measures the environment $E'$ while keeping
intact the entanglement of $Q'$ with $R$. It was shown
in~\cite{bib_schum2} that $L_Q=0$ is in fact a necessary {\em and}
sufficient condition for this to be feasible.  Such a transfer of
entanglement corresponds to the quantum equivalent of error
correction, and will be discussed with reference to the quantum Fano
inequality in Section III.D.

\subsection{Axioms for quantum information}
In the following, we present a number of reasonable ``axioms'' for a 
quantum mutual information, and show that $I_Q$ defined above has the required
properties. These are:

\begin{itemize}
\item[(i)] non-negativity
\item[(ii)] concavity in $\rho_Q$ (for a fixed channel)
\item[(iii)] convexity in $\rho_Q'$ (for fixed $\rho_Q$)
\item[(iv)] subadditivity
\end{itemize}

These requirements for a quantum mutual entropy (``entanglement
processed by the channel'') are very natural and reflect the kind of
requirements that are put on classical channels. 
The non-negativity of $I_Q$ is simply a consequence of the
subadditivity of quantum entropies. (Just like the mutual Shannon entropy,
the mutual quantum entropy is a non-negative quantity). 
Concavity of quantum information in $\rho_Q$
[axiom (ii)] reflects that the information processed by a
channel with a mixture of quantum states $\rho_Q=\sum_i w_i\rho_Q^i$ 
(with $\sum_i w_i=1$) as
input should be larger than the average information processed by
channels that each have a mixture $\rho_Q^i$ as input, i.e.,
\be
I_Q(\rho_Q)\geq\sum_i w_i I_Q(\rho_Q^i)\;.
\ee 
This is the quantum analogue of the concavity of the Shannon mutual 
information $H(X{\rm:}Y)$ in the input probability distribution $p(x)$
for a fixed channel, i.e., fixed $p(y|x)$. The proof uses
that, if the quantum operation achieved by the channel
is fixed, we have
\begin{eqnarray}
\rho'_{QE} &=& U_{QE} \left( \sum_i w_i \rho^i \otimes 
                             |0\rangle\langle 0|\right)
               U_{QE}^{\dagger}  \nonumber\\
           &=& \sum_i w_i U_{QE}(\rho^i \otimes |0\rangle\langle 0|)
                U_{QE}^{\dagger}  \nonumber\\ 
           &=& \sum_i w_i \rho'^i_{QE}\;.
\end{eqnarray}
Therefore, using
\begin{eqnarray}
I_Q(\rho_Q) &=& S(R{\rm:}Q') \nonumber\\
            &=& S(R)+S(Q')-S(RQ') \nonumber\\
            &=& S(Q'E')+S(Q')-S(E') \nonumber\\
            &=& S(Q'|E')+S(Q')
\label{eq-24}
\end{eqnarray}
the concavity of the quantum information in the input
results from the concavity of $S(Q'|E')$ in $\rho'_{QE}$
and from the concavity of $S(Q')$ in $\rho'_Q$~\cite{bib_wehrl}.
\par

Convexity of the processed information in $\rho_Q'$ [axiom
(iii)] states that, if the superoperator that takes 
a fixed $\rho_Q$ into $\rho_Q'$ is such that
\be
\rho_Q'=\sum_j w_j \rho'^j_Q\;,
\ee
then
\be 
I_Q(\rho_Q\to \rho_Q') \leq \sum_j w_j I_Q(\rho_Q\to \rho'^j_Q)\;.
\ee
Thus, the processed information of a channel that is a
``superposition'' of
channels (each used with probability $w_j$)
that result in $\rho_Q'$ cannot exceed the average of the
information for each channel. One has a similar property
for classical channels: the mutual information $H(X{\rm:}Y)$
is a convex function of $p(y|x)$ for a fixed input distribution $p(x)$.
The proof follows from noting that, if the input is fixed,
we have
\be
\rho'_{RQ}=\sum_j w_j \rho'^j_{RQ}
\ee
Then, expressing the quantum information as
\be
I_Q(\rho_Q\to \rho_Q') = S(R{\rm:}Q') = S(R)-S(R|Q')\;.
\ee
and noting that $S(R)$ is constant, the concavity of
$S(R|Q')$ in $\rho'_{RQ}$ implies the convexity of the quantum
information in the output.

\par
\begin{figure}
\caption{Parallel channels as quantum network, in the derivation of
  the subadditivity of mutual von Neumann entropies. The
  entanglement between $Q_1$, $Q_2$, and the reference is indicated by a 
  dashed line.}
\vskip 0.25cm
\centerline{\psfig{figure=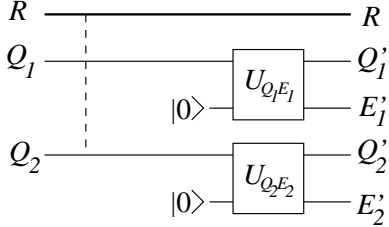,width=2.0in,angle=-90}}
\label{channel-figsub}
\vskip -0.25cm
\end{figure}

Finally, the subadditivity of quantum information [axiom (iv)] is a
condition which ensures that the information processed by a joint
channel with input $\rho_{Q_1Q_2}$ is smaller or equal to the
information processed ``in parallel'' by two channels with input
$\rho_{Q_1}=\tra_{Q_2}(\rho_{Q_1Q_2})$ and
$\rho_{Q_2}=\tra_{Q_1}(\rho_{Q_1Q_2})$ respectively.
Thus, if $R$ is the reference system purifying the joint input
$Q_1Q_2$, $Q_1$ is purified by $RQ_2$ while $Q_2$ is purified by
$RQ_1$ (see Fig.~\ref{channel-figsub}). 

The subadditivity of von Neumann mutual entropies for such a
channel can be written as
\be    \label{eq_subadditiv}
S(R{\rm:} Q_1' Q_2') \leq S(RQ_2{\rm:}Q_1')+S(RQ_1{\rm:}Q_2')\;,
\ee
which can be read as
\be
I_{12}\leq I_1 + I_2
\ee
with the corresponding identifications, and mirrors the classical
inequality
\be
H(X_1X_2{\rm:}Y_1Y_2)\leq H(X_1{\rm:}Y_1)+H(X_2{\rm:}Y_2)
\ee
for two independent channels taking $X_1\to Y_1$ and $X_2\to Y_2$.  

To prove inequality~(\ref{eq_subadditiv}), we
rewrite the quantum information of each channel
using Eq.~(\ref{eq-24}) and the fact that $E_1$ and $E_2$ are initially
in a {\em product} state.
Eq.~(\ref{eq_subadditiv}) then becomes
\be
&&S(Q_1'Q_2'|E_1'E_2')+S(Q_1'Q_2')\leq\nonumber \\
&&S(Q_1'|E_1')+S(Q_1')+S(Q_2'|E_2')+S(Q_2')\;.
\ee
Subadditivity of {\em conditional} entropies, i.e.,
\be
&&\hspace{-0.3cm}S(Q_1'Q_2'|E_1'E_2')\nonumber\\
&=&S(Q_1'|E_1'E_2')+S(Q_2'|E_1'E_2')-
\underbrace{S(Q_1'{\rm:}Q_2'|E_1'E_2')}_{\geq0}\nonumber\\
&\leq&S(Q_1'|E_1'E_2')+S(Q_2'|E_1'E_2')\nonumber\\
&\leq&S(Q_1'|E_1')-\underbrace{S(Q_1'{\rm:}E_2'|E_1')}_{\geq0}
+S(Q_2'|E_2')-\underbrace{S(Q_2'{\rm:}E_1'|E_2')}_{\geq0}\nonumber\\
&\leq& S(Q_1'|E_1')+ S(Q_2'|E_2')\;,
\ee
together with the subadditivity property of ordinary (marginal)
von Neumann entropies, proves Eq.~(\ref{eq_subadditiv}). The terms that
are ignored in the above inequality are positive due to strong
subadditivity. This property of subadditivity of the information
processed by quantum channels can be straightforwardly extended 
to $n$ channels. 

An alternative definition for the quantum information processed by a
channel, called ``coherent information'', has been proposed by
Schumacher and Nielsen~\cite{bib_schum2}, and by
Lloyd~\cite{bib_lloyd}.
This quantity $I_e=S(R'|E')=S-L_Q$ is not positive [axiom (i)], and
violates axioms (ii) and (iv), which leads to a {\em violation} of the
reverse data-processing inequality, while the
``forward'' one is respected~\cite{bib_schum2} (as opposed to the von
Neumann mutual entropy which observes both, see below).
The coherent information attempts to capture the ``purely'' quantum
piece of the processed information while separating out any classical 
components. This separation appears to be at the origin of the
shortcomings mentioned above. 

\subsection{Inequalities for quantum channels}
From the properties of the ``mutual entanglement'' $I_Q$ 
derived above, we can prove data-processing inequalities for
$I_Q$ which reflect probability conservation, as well as the Fano
inequality which relates the loss of a channel to the fidelity of a code. 

\begin{figure}
\caption{Chaining of channels in the derivation of the data-processing 
inequality. The output $Q_1$ is subjected to a second channel by entangling 
with an environment $E_2$ independent from $E_1$, to give output $Q_2$.}
\vskip 0.25cm
\centerline{\psfig{figure=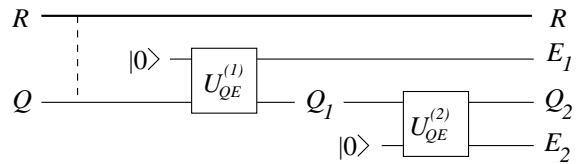,width=3in,angle=-90}}
\label{fig_dpi}
\vskip -0.25cm
\end{figure}

\vskip 0.25cm
\noindent{\it (i) Data-processing}
\vskip 0.25cm
Assume that starting with
the entangled state $QR$, entanglement with environment $E_1$ produces
the mixed state $Q_1$. This output is used again as an input to 
another channel, this time
entangling $Q_1$ with $E_2$ to obtain $Q_2$ (see Fig.~\ref{fig_dpi}).
The quantum analogue of the (forward) data-processing inequality
(\ref{dataproc}) that holds
for mutual informations in classical channels involves the mutual 
entanglements $S(R{\rm:}Q_1)$ and $S(R{\rm:}Q_2)$, and asserts that 
the mutual entanglement between reference and output cannot be increased by 
any further ``processing'':
\be
S(R{\rm:}Q_2)\leq S(R{\rm:}Q_1)\leq 2S\;.\label{qdatapr}
\ee
That such an inequality should hold is almost obvious from the 
definition of the mutual entanglement, but a short proof is given below. 
This proof essentially follows Ref.~\cite{bib_schum2},
and is based on the property of strong subadditivity applied to 
the system $RE_1E_2$:
\be
S(R{\rm:}E_2|E_1)=S(R{\rm:}E_1E_2)-S(R{\rm:}E_1)\geq 0\;.\label{strongsub}
\ee
For the channel $Q\rightarrow Q_1$, 
we see easily (see Fig.~\ref{fig_loss}) that
\be
S(R{\rm:}E_1) &=& S(R{\rm:}Q_1E_1)-S(R{\rm:}Q_1|E_1) \nonumber\\
& = & 2S-S(R{\rm:}Q_1)\;. \label{app1}
\ee
Similarly, considering $E_1E_2$ as the environment for the ``overall'' 
channel $Q\rightarrow Q_2$, we find
\be
S(R{\rm:}E_1E_2)=2S-S(R{\rm:}Q_2)\;. \label{app2}
\ee
Plugging Eqs.~(\ref{app1}) and (\ref{app2}) into the positivity condition 
(\ref{strongsub}), we obtain the quantum data processing inequality, 
Eq.~(\ref{qdatapr}), as claimed. 
\par

The {\em reverse} quantum data-processing inequality implies
that the entanglement processed by the second leg of the
channel, $S(RE_1{\rm:}Q_2)$, must be larger than the entanglement
processed by the entire channel:
\be\label{eq36}
S(R{\rm:}Q_2) \leq S(R E_1{\rm:}Q_2)\leq S(R E_1 E_2{\rm:}Q_2)= 2 S(Q_2)\;.
\ee
The proof relies on strong subadditivity applied to $Q_2 E_1 E_2$:
\be
S(Q_2{\rm:}E_1|E_2)=S(Q_2{\rm:}E_1E_2)-S(Q_2{\rm:}E_2)\geq 0\;.
\label{strongsub2}
\ee
For treating the channel $Q_1\rightarrow Q_2$ (i.e., the ``second
leg''), we have to purify 
the input state of $Q_1$, that is consider $RE_1$ as the ``reference''.
Thus, we have
\be
S(Q_2{\rm:}RE_1) = 2 S(Q_2) - S(Q_2{\rm:}E_2)\;.
\ee
For the ``overall'' channel $Q\rightarrow Q_2$, we have 
\be
S(Q_2{\rm:}R)=2 S(Q_2) - S(Q_2{\rm:}E_1 E_2)\;. 
\ee
These two last equations together with Eq.~(\ref{strongsub2}),
result in the reverse quantum data-processing inequality, Eq.~(\ref{eq36}).
\par

From Eq.~(\ref{qdatapr}) we obtain immediately an inequality relating
the loss of entanglement after the first stage $L_1$ (we drop the
index $Q$ that indicated the quantum nature of the loss in this
discussion), with the overall loss, 
$L_{12}$:
\be
0\leq L_1\leq L_{12}\;. \label{lossineq}
\ee
Physically, this implies that the loss $L_{12}$ cannot decrease from
simply chaining channels, just as in the classical case.  As
emphasized earlier, the loss $L_1$ corresponds to the share of initial
entanglement that is irretrievably lost to the environment. Indeed, if
the environment cannot be accessed (which is implicit by calling it an
environment) the decoherence induced by the channel cannot be
reversed. Only if $L_1=0$ can this be achieved~\cite{bib_schum2}. 
In view of this fact, it is natural to seek for a quantum
equivalent to the classical Fano inequality~(\ref{fanocl}).

\vskip 0.25cm
\noindent{\it (ii) Fano inequality}
\vskip 0.25cm

To investigate this issue, let us consider the chained channel  
above, where error correction has taken place via transfer of entanglement
with a second environment. Let us also recall the definition of 
``entanglement fidelity'' of Schumacher~\cite{bib_schum1}, which is 
a measure of how faithfully the dynamics of the channel has preserved 
the initial entangled quantum state $QR$:
\be
F_e(QR,Q'R) = \la QR|\,\rho_{Q'R}\,|QR\ra\equiv F_e^{QQ'}\;.\label{fid}
\ee
Since this entanglement fidelity does not depend on the reference 
system~\cite{bib_schum1}, we drop $R$ from $F_e$ from here on, as indicated 
in Eq.~(\ref{fid}).

Naturally, the entanglement fidelity can be related to the
probability of error of the channel.  The quantum analogue of the
classical Fano inequality should relate the fidelity of the {\em code}
(in our example above the fidelity between $QR$ and $Q_2R$, the
error-corrected system) to the loss of the error-correcting channel $L_{12}$.
The derivation of such an inequality is immediate
using the Fano-type inequality derived by Schumacher~\cite{bib_schum1}, 
which relates
the entropy of the environment of a channel $S(E')$
to the fidelity of entanglement,
\be
S(E')\leq H_2[F_e^{QQ'}]\,+\,(1-F_e^{QQ'})\log{(d_Qd_R-1)}\;,\label{eqfano}
\ee
where $d_Q$ and $d_R$ are the Hilbert-space dimensions of $Q$ and $R$ 
respectively, and $H_2[F]$ is again the dyadic Shannon entropy. 
Let us apply this
inequality to an error-correcting channel (decoherence +
error-correction), i.e., 
the chained channel considered above. In that case, the environment is 
$E_1E_2$, and the entanglement fidelity is now between $Q$ and $Q_2$, i.e.,
the fidelity of the {\em code}, and we obtain
\be
S(E_1E_2)\leq H_2[F_e^{QQ_2}]\,+\,(1-F_e^{QQ_2})\log{(d-1)}\;. 
\ee
Here, $d=d_R\,d_{Q_2}$ can be viewed as the Hilbert space dimension of
the code (this is more apparent in superdense coding discussed in the
Section VI).
To derive  the required relationship, we simply note that 
\be
S(E_1E_2) \geq L_{12}/2 
\ee
[this is Eq.~(\ref{ineq2}) applied to the composite channel]. This relates
the fidelity of the code $F_e^{QQ_2}$ to the loss $L_{12}$, yielding the Fano 
inequality for a quantum code
\be
L_{12}\leq 2\left[H_2[F_e^{QQ_2}]+
\left(1-F_e^{QQ_2}\right)\log{(d-1)}\right]\;. \label{fano}
\ee
As we noticed throughout the construction of quantum channels, a factor 
of 2 appears also in the quantum Fano inequality, commensurate with 
the fact that the loss can be twice the initial entropy. 
Inequality (\ref{fano}) puts an upper limit on the fidelity 
of a code for any non-vanishing loss $L_{12}$.

\section{Classical use of quantum channel}
In recent papers~\cite{bib_hausladen,bib_kholevo97}, the capacity for
the transmission of {\em classical} information through quantum channels has
been discussed. Essentially, this capacity is equal to the
maximal accessible information $\chi$ in the system, known as the Kholevo
bound~\cite{bib_kholevo}. 

\begin{figure}
\caption{Entanglement between $Q$, $R$, and the ancilla (or preparer)
  $X$ after measurement of the initial state of $Q$ by $X$, but prior
  to entanglement with the environment. The initial state of $Q$
  (before decoherence) is kept in memory, as it were, by $X$ via
  classical correlation with $Q$. }
\vskip 0.25cm
\centerline{\psfig{figure=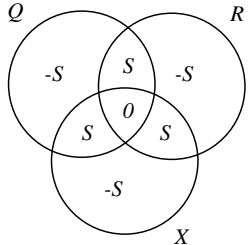,width=1.25in,angle=-90}}
\label{fig_trip}
\vskip -0.25cm
\end{figure}

\begin{figure}
\caption{Unitary transformation entangling the ``preparer'' (or
  alternatively, the classical ``memory'') $X$ with the pure
  environment $E$ and the quantum system $Q$. Neither the reference
  $R$ nor the preparer $X$ are affected by this operation. As the
  ternary Venn diagram between $Q'$, $E'$ and $X'$ is not pure in this
  case, mutual entropy between $Q'$ and $X'$ {\em can} be shared by
  $E'$. 
\label{classic-fig1}}
\vskip 0.25cm
\centerline{\psfig{figure=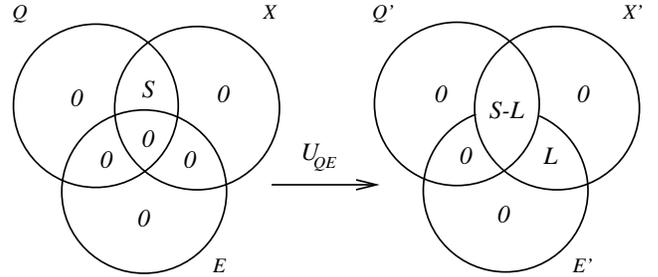,width=3.3in,angle=-90}}
\vskip -0.25cm
\end{figure}

What we show in the following is that the mutual entanglement
introduced in the previous section, i.e., the quantum mutual entropy
$S(R:Q')$ between the ``decohered'' quantum state $Q'$ and the
``reference'' state $R$, reduces to $\chi$ if the quantum state is
measured before it is transmitted, or, equivalently, if Q is prepared
by a classical ``preparer'' $X$. Let the system $QR$ be ``purified'' again
via a Schmidt decomposition as in Eq.~(\ref{eq9}). If we measure
$Q$ in its eigenbasis we can write
\be
|RXQ\ra=\sum_i \sqrt{p_i}\,|r_i\,x_i\, i\ra\;,
\ee
where $x_i$ are the eigenstates of $X$ (if $X$ is in state $x_i$, $Q$
is in state $i$ etc.). (Figure~\ref{fig_trip} summarizes the
relationship between the respective entropies.)
Naturally then, tracing over $R$ we obtain
\be
\rho_{XQ}=\sum_ip_i\,|x_i\ra\la x_i|\otimes \rho_i\label{eq28}
\ee
with $\rho_i=|i\ra\la i|$, and similarly for $\rho_{RQ}$. 
Thus, $X$ and $Q$ are {\em classically}
correlated: each state of the ``preparer'' $X$ represents a state of
$Q$, or alternatively, $X$ reflects (keeps in memory) the initial quantum
state of $Q$. If the entropy of the quantum system $Q$ before
transmission is $S$ (just like in the previous section), the mutual
entropy between $R$ and $Q$ (as well as between $X$ and $Q$) is also
$S$, unlike the value $2S$ found in the quantum use. Decoherence now
affects $Q$ by entangling it with the environment, just like earlier. 
Thus, 
\be
\rho_{XQ}\to\rho_{XQ'}=\sum_ip_i|x_i\ra\la x_i|\otimes \rho'_i
\ee
where
\be
\rho_i^\prime=\tra_E\left\{U_{QE}\,\left(\rho_i\otimes
  |0\ra\la0|\right)\,U^\dagger_{QE}\right\}\;, \label{eq31}
\ee
and we assumed again that the environment $E$ is in a fixed ``0'' state
before interacting with $Q$. Now our proof proceeds as before, only
that the loss in the ``classical'' channel obeys different
inequalities. The requirement that the entangling
operation $U_{QE}$ does not affect $X$ or $R$ now implies
\be
S(X'{\rm:}E'Q')=S(X{\rm:}Q)=S(R{\rm:}Q)= S \label{eq32}
\ee 
(see Figure~\ref{classic-fig1}). 
Applying the chain rule to the left hand side of Eq.~(\ref{eq32}) leads to
\be
S(X'{\rm:}E'Q')= S(X{\rm:}Q')+S(X{\rm:}E'|Q')\;.\label{eq33}
\ee
The quantum mutual entropy between the preparer and the quantum state 
after decoherence, $S(X{\rm:}Q')$, can be shown to be equal to the Kholevo 
bound $\chi$ (see Ref.~\cite{bib_access}). With $L=S(X{\rm:}E'|Q')$ 
(the classical loss of the channel) we thus conclude from
Eqs.~(\ref{eq33}) and (\ref{eq32}) that
\be
S=\chi+L\;.
\ee
Note that $S(X{\rm:}Q')$ is equal to $S(R{\rm:}Q')$, the mutual 
entanglement $I_Q$ introduced earlier, as 
$S(X)=S(R)$ and $S(XQ')=S(RQ')$. 
Thus, 
\be
I_Q\equiv S(R:Q')=\chi
\ee
if known quantum states are sent through the
channel, as advertised. It was shown recently by Kholevo~\cite{bib_kholevo97}
that the maximum of the latter quantity indeed plays the role of
channel capacity for classical information transmission
\be
C=\max_{p_i}\,\left[S(\rho')-\sum_i p_iS(\rho'_i)\right]\equiv\max_{p_i}\,\chi 
\ee
where $\{p_i\}$ is a probability distribution of symbols at the
source, and $\rho'_i$ are the (not necessarily orthogonal) quantum
states received at the output, with the probability
distribution $\{p_i\}$ and $\rho'=\sum_i p_i\rho'_i$.
Thus, the quantity $C_Q$ that we propose as a capacity for
entanglement/correlation transmission reverts to the capacity for information
transmission $C$ if the unknown quantum states are {\em measured} before
transmission. This represents solid evidence in favor of our interpretation.

Let us now calculate
the quantities introduced here for a specific simple model of quantum noise. 

\section{Quantum depolarizing channel}
The quantum depolarizing channel is an idealization of a quantum
storage and transmission process in which the stored quantum state can
undergo bit-flip and phase errors. This is not the most general
one-qubit channel\footnote{A more general depolarizing channel could
  be constructed by allowing each of the possible errors a different
  probability.}, but appears to be sufficient to examine a number of
interesting aspects of quantum communication.
\subsection{Quantum use}
Imagine a quantum state 
\be |\Psi\rangle =
\alpha\,|0\rangle + \beta\,|1\rangle\;, 
\ee 
where the basis states of
the qubit can be taken to be spin-1/2 states polarized in the
$z$-direction, for example. (Specifically, we use the convention
$\sigma_z|1\rangle=|1\rangle$.)  The depolarizing channel is
constructed in such a way that, due to an interaction with an
environment, the quantum state survives with probability $1-p$, but is
depolarized with probability $p/3$ by either a pure bit-flip, a pure
phase-error, or a combination of both: 
\be 
|\Psi\ra &\stackrel{1-p}\longrightarrow & |\Psi\ra\;,\nonumber\\ 
|\Psi\ra& \stackrel{p/3}\longrightarrow & \sigma_x|\Psi\ra=
\alpha\,|1\ra+\beta\,|0\ra\;, \nonumber\\ 
|\Psi\ra &\stackrel{p/3}\longrightarrow & \sigma_z|\Psi\ra=
-\alpha\,|0\ra+\beta\,|1\ra\;, \nonumber\\ 
|\Psi\ra &\stackrel{p/3}\longrightarrow & \sigma_x\sigma_z|\Psi\ra=-
\alpha\,|1\ra+\beta\,|0\ra \;, 
\ee 
where the $\sigma$ are Pauli matrices. Such an ``arbitrary'' quantum
state $\Psi$ can, without loss of generality, considered to be a state
$Q$ that is entangled with a reference state $R$, such that
the marginal density matrix of $Q$ can be written as 
\be 
\rho_Q =q\,|0\ra\la0|\,+\,(1-q)\,|1\ra\la1|
\label{rhomix} 
\ee 
with entropy $S(\rho_Q)=-\tra\rho_Q\log\rho_Q=H_2[q]$ and $q$ a probability
$(0\leq q\leq1$). In other words, the coefficients $\alpha$ and
$\beta$ need not be complex numbers. Conversely,
we can start with such a mixed state at the input, and consider $QR$
as a {\em pure} quantum state that this mixed
state obtains from. For example, 
\be 
|QR\ra = \sqq\,|10\ra\, - \,\sq\,|01\ra\;.  \label{qr}
\ee 
Naturally then, the mixed state Eq.~(\ref{rhomix}) is obtained by simply
tracing over this reference state. Pure states with real coefficients
such as (\ref{qr}) are not general, but suffice for the depolarizing channel
as $R$ is always traced over.

Let us now construct a basis for $QR$ that interpolates between
completely independent and completely entangled states, and allows us
to choose the initial entropy of $Q$ with a single parameter $q$. We thus
introduce the orthonormal ``$q$-basis'' states 
\be 
|\Phi^-(q)\ra &=& \sqq\,|00\ra\,-\,\sq\,|11\ra\;, \nonumber \\ 
|\Phi^+(q)\ra &=& \sq\,|00\ra\,+\,\sqq\,|11\ra\;, \nonumber \\ 
|\Psi^-(q)\ra &=& \sqq\,|10\ra\,-\,\sq\,|01\ra\;, \nonumber \\ 
|\Psi^+(q)\ra &=& \sq\,|10\ra\,+\,\sqq\,|01\ra\;.  
\ee 
Note that for $q=0$ or 1, these states are product states,
while for $q=1/2$ they are completely entangled, and $\Psi^\pm(1/2)$
and $\Phi^\pm(1/2)$ are just the usual Bell basis states. The
possibility of quantum decoherence of these states is introduced by
entangling them with an environment in a pure state, taken to be of
the same Hilbert space dimension as $QR$ for simplicity, i.e., a
four-dimensional space for the case at hand. This is the
minimal realization of a depolarizing channel. 

\begin{figure}
\caption{Mutual entanglement between the depolarized state $Q'$ and the 
reference system $R'=R$, as a function of error $p$ and  
parameter $q$. Note that the channel is 100\% depolarizing
at $p=3/4$. The concavity in $q$ [according to axiom (ii)] as well as
the convexity in $p$ [axiom (iii)] are apparent.}
\vskip 0.25cm
\centerline{\psfig{figure=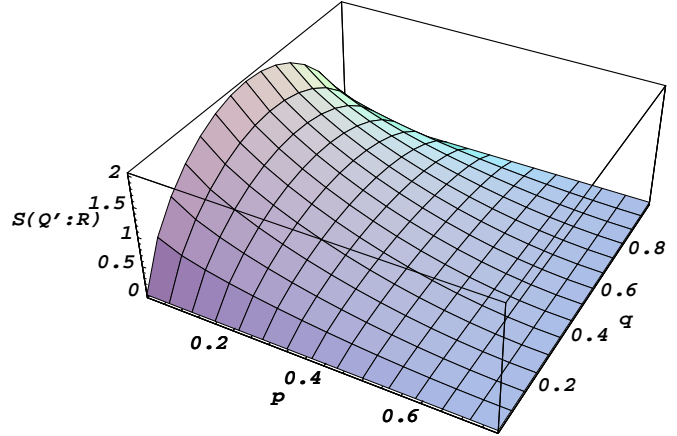,width=3.5in,angle=0}}
\label{fig_3d}
\vskip -0.25cm
\end{figure}

Let us assume that $QR$
(for definiteness) is initially in the state $|\Psi^-(q)\ra$, and the
environment in a superposition
\be 
|E\ra &= &\sqrt{1-p}\,|\Psi^-(q)\ra \nonumber \\
&+&\sqrt{p/3}\left(|\Phi^-(q)\ra+|\Phi^+(q)\ra+|\Psi^+(q)\ra\right)\;.  
\ee 
The environment and $QR$ are then entangled by means of the unitary operator 
$U_{QRE}=U_{QE}\otimes 1_R$, with 
\be 
U_{QE} &=
&1\otimes P_{\Psi^-}(q)+
\sigma_x\otimes P_{\Phi^-}(q)\nonumber\\
&+&(-i\sigma_y)\otimes P_{\Phi^+}(q)+ 
\sigma_z\otimes P_{\Psi^+}(q)\;,
\ee 
where the $P_{\Phi}(q)$ and $P_{\Psi}(q)$ stand for projectors
projecting onto $q$-basis states. Note that the Pauli matrices act
only on the first bit of the $q$-basis states, i.e., the entanglement
operation only involves $Q$ and $E$. Depending on the entanglement
between $Q$ and $R$, however, this operation also affects the
entanglement between $R$ and $E$. Thus, we obtain the state 
\be
\lefteqn{|Q^\prime R^\prime E^\prime\ra = U_{QRE}|QR\ra|E\ra = }\nonumber \\
&& \sqrt{1-p}\,|\Psi^-_{QR}(q),\,\Psi^-_E(q)\ra +
\sqrt{p/3}\left(|\Phi^-_{QR}(q)\;,\Phi^-_E(q)\ra +\right.\nonumber\\
&&\left.|\Phi^+_{QR}(1-q)\;,\Phi^+_E(q)\ra +
 |\Psi^+_{QR}(1-q)\;,\Psi^+_E(q)\ra\right)
\ee
on account of the relations 
\be 
\sigma_x |\Psi^-_{QR}(q)\ra & = &
|\Phi^-_{QR}(q)\ra \;, \\ (-i\,\sigma_y) |\Psi^-_{QR}(q)\ra & = &
|\Phi^+_{QR}(1-q)\ra \;,\\ \sigma_z |\Psi^-_{QR}(q)\ra & = & 
|\Psi^+_{QR}(1-q)\ra\;,
\ee
and with obvious notation to distinguish the environment ($E$) and
quantum  system ($QR$) basis states. The (partially depolarized) 
density matrix 
for the quantum system is obtained by tracing over the environment:
\be
\rho_{Q'R'} & = & \tra_E\left(|Q'R'E'\ra\la Q'R'E'|\right) = 
(1-p)\,P_{\Psi^-}(q) +\nonumber \\
& & p/3\left[P_{\Phi^-}(q)+P_{\Phi^+}(1-q)+P_{\Psi^+}(1-q)\right]\;.
\ee
Its eigenvalues can be obtained to calculate the entropy:
\be
S_e(p,q)\equiv S(Q'R') = \nonumber \hspace{4.5cm}\\
H[\frac{2p}3(1-q),\frac{2pq}3,\frac12(1-\frac{2p}3+\Delta),
\frac12(1-\frac{2p}3-\Delta)]\;,
\ee
with $H[p_1,...,p4]$ the Shannon entropy, and 
\be
\Delta = \left[(1-2p/3)^2-16/3\,p(1-p)\,q(1-q)\right]^{1/2}\;.
\ee
By tracing over the reference state we obtain the density matrix of 
the quantum system after the interaction $\rho_{Q'}$, and its respective 
entropy
\be
S'(p,q)\equiv S(Q')= H_2[q+\frac{2p}3(1-2q)]\;.
\ee
Together with the entropy of the reference state (which is unchanged 
since $R$ was not touched by the interaction), $S(R')=S(R)=H_2[q]$, 
this is enough to fill in the ternary entropy diagram reflecting the 
dynamics of the channel, Fig.~\ref{fig_loss}. We thus find the 
mutual entanglement processed by the channel:
\be
I_Q=S(Q'{\rm:}R)=2H_2[q]-L_Q(p,q)\;, 
\ee
where the loss is 
\be
L_Q(p,q)=H_2[q]-H_2[q+\frac{2p}3(1-2q)]+ S_e(p,q)\;.
\ee
The mutual entanglement is plotted in Fig.~\ref{fig_3d}, 
as a function of the error probability $p$ of the channel 
and of the parameter $q$ which determines the initial entropy.

\begin{figure}
\caption{Maximal entanglement transfer $C(p)$ and minimal loss $L(p)$
as a function of the error probability $p$.}
\vskip 0.25cm
\centerline{\psfig{figure=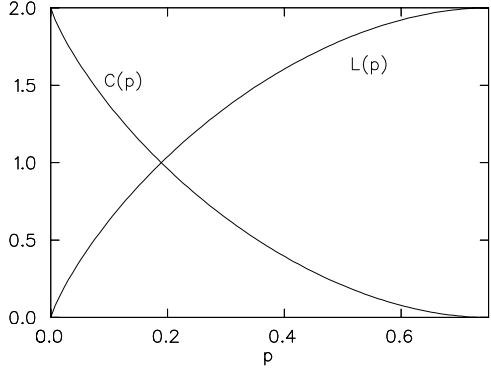,width=2.5in,angle=90}}
\label{fig_depol}
\vskip -0.25cm
\end{figure}

The mutual entanglement is maximal when the entropy of the 
source is maximal (as in the classical theory), i.e., $q=1/2$. Then:
\be 
C_Q &=& \max_q\, I_Q\nonumber\\
 &=& 2-S_e(p,1/2) 
  = 2-H_2[p]-p\,\log3\;. \label{depolcap}
\ee
In that case, the maximal rate of entanglement transfer is 2 bits
(error-free transfer, $p=0$). The capacity only vanishes at $p=3/4$,
i.e., the 100\% depolarizing channel. This is analogous to the
vanishing of the classical capacity of the binary symmetric channel at
$p=1/2$.  As an example of such a channel, we shall discuss the
transmission of the entanglement present in a Bell state (one out of
four fully entangled qubit pairs) through a ``superdense coding''
channel in Section VI.A.  The maximal mutual entanglement and minimal
loss implied by Eq.~(\ref{depolcap}) are plotted in
Fig.~\ref{fig_depol} as a function of $p$.  This error rate $p$ can be
related to the fidelity of the channel by
\be
F_e^{Q'Q}= 1-p+\frac p3\,(1-2q)^2\;,
\ee
where $F_e^{Q'Q}$ is Schumacher's fidelity of entanglement introduced earlier.
Note that this implies that the Fano inequality Eq.~(\ref{eqfano}) is
saturated at $q=1/2$ for any $p$. 
\subsection{Classical use}
Now, instead of using the channel to transmit entanglement (sending
unknown quantum states), one could equally 
well use it to send classical information (known quantum states) as
outlined in section IV. Here, we calculate the capacity for the
transmission of classical information through the quantum depolarizing
channel and verify that the result is equal to the value obtained by
Calderbank and Shor~\cite{bib_calder} using the Kholevo theorem. 

Before entanglement with the environment, let us then measure the mixed state 
$Q$ via an ancilla $X$, after which $Q$ and $X$ are classically correlated, 
with mutual entropy $H_2[q]$. 
Note that this operation leads to an entangled triplet $QRX$ 
at the outset, as in Fig.~\ref{fig_trip}, with $S=H_2[q]$. 
We now proceed with the calculation
as before. The basis states for the system $|QXR\ra$ are then simply
\be
|\Phi_X^-(q)\ra &=& \sqq\,|000\ra\,-\,\sq\,|111\ra\;, \nonumber \\ 
|\Phi^+_X(q)\ra &=& \sq\,|000\ra\,+\,\sqq\,|111\ra\;, \nonumber \\ 
|\Psi^-_X(q)\ra &=& \sqq\,|110\ra\,-\,\sq\,|001\ra\;, \nonumber \\ 
|\Psi^+_X(q)\ra &=& \sq\,|110\ra\,+\,\sqq\,|001\ra\;, 
\ee
where we used the index $X$ on the basis states to distinguish them
from the two-qubit basis states introduced earlier. The entanglement operation 
is as before, with a unitary operator acting on $Q$ and $E$ only. 
Because of the additional trace over the ancilla $X$, however, 
we now find for the density matrix $\rho_{Q'R'}$:
\be
\rho_{Q'R'} 
& = & (1-2p/3)\left[\,(1-q)|10\ra\la10|\,+\,q|01\ra\la 01|\,\right]\nonumber\\ 
&+&2p/3\left[\,(1-q)|00\ra\la00|\,+\,q|11\ra\la11|\,\right]\;.
\ee
Consequently, we find for the mutual information transmitted through the
channel 
\be
I = S(Q'{\rm:}R)=H_2[q]-L(p,q)\;,
\ee 
with the (classical) loss of information 
\be
L(p,q) &=&H[\frac{2p}3(1-q),\frac{2p}3q,
(1-\frac{2p}3)(1-q),(1-\frac{2p}3)q]\nonumber\\
 &-&H_2[q+\frac{2p}3(1-2q)] \;.
\ee
Maximizing over the input distribution as before, we obtain
\be
C= \max_q S(Q'{\rm:}R) = 1-H_2[2p/3]\;, \label{classcap}
\ee
the result derived recently for the depolarizing channel simply from
using the Kholevo theorem~\cite{bib_calder}.  Note that
Eq.~(\ref{classcap}) is just the Shannon capacity of a binary symmetric
channel~\cite{bib_ash}, with a bit-flip probability of $2p/3$ (of the
three quantum error ``syndromes'', only two are classically detectable
as bit-flips).

\section{Interpretation}
\subsection{Quantum capacity and superdense coding}
\begin{figure}
\caption{Quantum Venn diagram for the noisy superdense coding
  channel before decoherence. Conditionally on the classical bits $C$, 
  $QR$ is in a pure entangled state described by a Venn diagram of the
  form $(-S,2S,-S)$. Note that no information about $C$ is contained
  in $R$ or $Q$ {\em alone}, i.e., $S(C{\rm:}R)=S(C{\rm:}Q)=0$. 
\label{fig-super}}
\vskip 0.25cm
\centerline{\psfig{figure=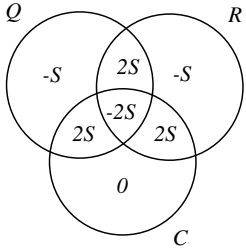,width=1.25in,angle=-90}}
\vskip -0.25cm
\end{figure}

The interpretation of the capacity suggested here as a quantum
mechanical extension of the classical construction
can be illustrated in an
intuitive manner with the example of the depolarizing channel
introduced above.  The idea is that $I_Q$ reflects the capacity for
transmission of quantum mutual entropy (entanglement and/or classical
information) but that the amount transferred in a particular channel
depends on how this channel is used. 
A particularly elegant channel
that uses $I_Q$ to its full extent is the noisy ``superdense coding''
channel. There, the entanglement between sender and receiver  is used
to transmit two bits of classical information by sending just {\em
  one} quantum bit~\cite{bib_superdense,bib_neginfo}. In a general
superdense coding scheme, the initial state $QR$ is one of a set of
entangled states conditionally on classical bits $C$. 
This situation
can be related to our previous discussion by noting that all
entropies appearing there are to be understood as
{\em conditional} on the classical bits $C$ that 
are to be sent through the channel as shown in Fig.~\ref{fig-super}. The
von Neumann capacity introduced above is then just
\be
I_Q=S(R:Q'|C)\;. \label{eq-81}
\ee
It is not immediately obvious that this von Neumann capacity is equal
to the {\em classical} capacity between preparer (usually termed
Alice) and the receiver (Bob). However, it is not difficult to
prove [using the fact that $S(R{\rm:}Q)=S(R{\rm:}C)=S(Q{\rm:}C)=0$] that
Eq.~(\ref{eq-81}) is in fact equal to the maximal
amount of classical information about $C$ extractable from $RQ'$
(after $Q$ decohered),
which is\footnote{That the quantum mutual entropy between a preparer and
  a quantum system is an upper bound to the amount of 
  classical  information obtainable by measuring the quantum system 
  (the Kholevo bound) is shown in Ref.~\cite{bib_access}.}
\be
\chi=S(RQ':C)\;.
\ee 
Thus, in this example the amount of entanglement processed in a channel
can be viewed as the amount of {\em classical} information about the
``preparer'' of the entangled state $QR$. This amount of information
can reach {\em twice} the entropy of $Q$ (2 bits in standard
superdense coding), which is classically impossible.
(The superdense coding and
teleportation channels will be discussed in detail elsewhere).
Having established this relation between superdense coding and the
general quantum channels treated here, let us imagine that the qubit
that is sent through the channel (and which is ``loaded'' with
entanglement) is subject to the depolarizing noise of the previous
section. Indeed, if $p=0$ the two classical bits can be decoded
perfectly, achieving the value of the capacity. It has been argued
recently~\cite{bib_neginfo} that this can be understood by realizing
that besides the qubit that is sent forwards in time in the channel,
the entanglement between sender and receiver can be viewed as an
antiqubit sent {\em backwards} in time (which is equivalent to a qubit
sent forwards in time if the appropriate operations are performed on
it in the future). Thus, the quantum mechanics of superdense coding
allows for the time-delayed (error-free) transmission of information,
which shows up as excessive capacity of the respective channel. On the
other hand, it is known that (for un-encoded qubits) superdense coding
becomes impossible if $p\approx0.189$, which happens to be the precise
point at which $I_Q=1$. This is related to the fact that at this point
the ``purification'' of ``noisy'' pairs becomes impossible.
However, the capacity of this channel is not zero. While no
information can be retrieved ``from the past'' in this case, the
single qubit that is sent through the channel still carries
information, indeed, it shares one bit of mutual entropy with the qubit
stored by the receiver.  Clearly, this is still a quantum channel: if
it were classical, the transmission of one bit could not take place
with unit rate and perfect reliability, due to the noise level
$p=0.189$.  As the receiver possesses both this particle and the one
that was shared earlier, he can perform joint measurements (in the
space $Q'R$) to retrieve at least one of the two classical bits.

An extreme example is the
``dephasing'' channel, which is a depolarizing channel with only
$\sigma_z$-type errors, affecting the phase of the qubit. 
As is well known, classical
bits are unaffected by this type of noise, while quantum
superpositions are ``dephased''. The channel becomes useless (for the
storage of superpositions) at $p=0.5$, yet measuring the qubit yields
one {\em classical} bit in an error-free manner. A calculation of 
$\max_q S(R:Q')$ for this channel indeed yields
\be
I_Q(p)=2-H_2[p]\;.
\ee
In this limiting case thus, it appears possible to separate the
classical ($I=1$) from the purely quantum capacity. However, it might
well be possible that this cannot be achieved in general.  Below, we
show that such an ``excessive'' von Neumann capacity (as in superdense
coding) is consistent with a commensurate quantum Hamming bound.

\subsection{Quantum Hamming bounds}
Classically, the Hamming
bound~\cite{bib_ash} is an upper bound on the number $s$ of codewords
(bit-strings of length $n$) for a code to correct $t$ errors:
\be
s\,\sum_{i=0}^t {n \choose i}\le 2^n\;. \label{classham}
\ee 
This is a necessary (but not sufficient) condition for error-free
coding, which reflects the necessary space to accommodate all the
codewords and associated descendants for all error syndromes.
For $s$ codewords coding for $k$ bits ($s=2^k$), we can
consider the asymptotics of (\ref{classham}) in the limit of
infinitely long messages ($n\rightarrow\infty$), and find that the
rate of error-free transmission is limited by
\be
R\le - \frac1n\log \sum_{i=0}^{pn}{n \choose i}
\left(\frac12\right)^i\left(\frac12\right)^{n-i} 
\ee
where $R=k/n$ is the transmission rate and $p=t/n$ is the asymptotic
probability of error. 
Using
\be
\lim_{n\to\infty} &-&\frac1n\log\left\{\sum_{i=0}^{pn} {n \choose i}
\,r^i\,(1-r)^{n-i}\right\} \nonumber \\
&=& p\,\log\frac pr\, + \, (1-p)\,\log\frac{1-p}{1-r}\nonumber\\
&\equiv&H(p,1-p\,\|\,r,1-r)\;,
\ee
where $H(p,1-p\,\|\,r,1-r)$ is the {\em relative} entropy between the
probability distributions $p$ and $r$, we can write
\be
R\le H(p,1-p\,\|\,1/2,1/2)=1-H_2(p)\;.
\ee
The relative entropy thus turns out to be just the classical capacity
of the channel, and measures the ``distance'' 
of the error-probability
of the channel relative to the ``worst case'', i.e., $p=1/2$. Note
that relative entropies are positive semi-definite. 

For quantum channels, the standard quantum Hamming bound for
non-degenerate (orthogonal) codes is written 
as~\cite{bib_laflamme,bib_ekert,bib_bdsw}
\be
2^k\,\sum_{i=0}^t 3^i{n \choose i}\le 2^n\;,
\ee
which expresses that the number of orthogonal states identifying the
error syndromes on the $2^k$ different messages must be smaller than 
$2^n$, the dimension of the Hilbert space of the quantum
state $Q$ ($n$ qubits). In the limit of large $n$, this translates
into an upper bound for the rate of non-degenerate quantum codes
\be 
R\le -\frac1n\log\left\{ \sum_{i=0}^{pn}{n \choose i}\left(\frac34\right)^i
\left(\frac14\right)^{n-i}  \right\} -1 \;.
\ee
which can (as in the classical case) be written in terms of a relative
entropy
\be
R\le H(p,1-p\,\|\,3/4,1/4)\,-\,1\,=\,1-S_e(p)\;,\label{usualqhb}
\ee
Thus, the usual quantum Hamming bound limits the rate of
non-degenerate quantum codes by
the capacity based on ``coherent information'' proposed 
in~\cite{bib_schum2,bib_lloyd}, which is thought of as the ``purely quantum''
piece of the capacity.
Note that the positivity of
relative entropy does {\em not} in this case guarantee such a capacity
to be positive, which may just be a reflection of the
``inseparability'' of the von Neumann capacity.

The quantum Hamming bound shown above relies on coding the error
syndromes only into the quantum state $Q$ that is processed, or, in
the case of superdense coding, sent through the noisy channel.  As we
noted earlier, however, a quantum system that is entangled does not,
as a matter of principle, have a state on its own. Thus, the entangled
reference system $R$ {\em necessarily} becomes part of the quantum
system, even if it is not subject to decoherence. Thus, the Hilbert
space available for ``coding'' automatically becomes as large as $2n$,
the combined Hilbert space of $Q$ and $R$. This is most obvious again
in superdense coding, where the ``decoding'' of the information
explicitly involves joint measurements of the decohered $Q'$ {\em and}
the ``reference'' $R$, shared between sender and receiver (in a
noise-free manner). 
The corresponding {\em
  entanglement} quantum Hamming bound therefore can be written by
remarking that while the coding space is $2n$, only $n$ qubits are
sent through the channel, and thus
\be
2^k\,\sum_{i=0}^t 3^i{n \choose i}\le 2^{2n}\;.
\ee
Proceeding as before, the rate of such quantum codes is limited by
\be
R\le H(p,1-p\,\|\,3/4,1/4)\,=\,2-S_e(p)\;, \label{entham}
\ee 
the von Neumann capacity $C_Q$ for the depolarizing channel 
proposed in this paper, Eqs.~(\ref{quantcap}) and (\ref{depolcap}). 
The latter is always positive, and represents the
``distance'' between the error probability $p$ of the channel and the
worst-case error $p=3/4$ (corresponding to a 100\% depolarizing
channel), in perfect analogy with the classical
construction. Eq.~(\ref{entham}) thus guarantees the {\em weak
converse} of the quantum fundamental theorem: that no code can be 
constructed that maintains a rate larger than the capacity $C_Q$ with a 
fidelity arbitrarily close to one.   
\section{Conclusions}
We have shown that the classical concept of information transmission
capacity can be extended to the quantum regime by defining a von
Neumann capacity as the maximum mutual von Neumann entropy between the
decohered quantum system and its reference. This mutual von Neumann
entropy, that describes the amount of information---classical and/or
quantum---processed by the channel, obeys ``axioms'' that any measure
of information should conform to. As for any quantum extension, the
von Neumann capacity reverts to its classical counterpart when the
information is ``classicized'' (i.e., it reverts to the Kholevo
capacity when measured or prepared states are sent), and ultimately to
the Shannon capacity if all quantum aspects of the channel are ignored
(i.e., if orthogonal states are sent and measured).  Thus, the von
Neumann capacity of a channel can only vanish when the classical
capacity is also zero, but it can be excessive as entanglement allows
for superdense coding.  In order to take advantage of this, however,
both the quantum system that decoheres {\em and} the reference system
it is entangled with need to be accessible.  In practical quantum
channels this appears to be impossible, and the rate of practical codes must
then be considerably smaller than the von Neumann capacity. Yet,
because of the inseparability of entangled states, a consistent
definition of channel capacity {\em has} to take into account the full
Hilbert space of the state.  Whether a capacity can be
defined {\em consistently} that characterizes the ``purely'' quantum
component of a channel is still an open question.

\acknowledgements 
We would like to thank John Preskill and the members of the QUIC group
at Caltech for discussions on the depolarizing channel, as well as
Howard Barnum and Michael Nielsen for discussions during the Quantum
Computation and Quantum Coherence Program at the ITP in Santa Barbara,
where most of this work was done.  This research was supported in part
by NSF Grant Nos. PHY 94-12818 and PHY 94-20470 at the Kellogg
Radiation Laboratory, and Grant No. PHY 94-07194 at the ITP in Santa
Barbara.

\end{multicols}

\begin{references}

\bibitem{bib_reviews} For reviews, see D. P. DiVincenzo, Science {\bf
270}, 255 (1995); S. Lloyd, Sci. Am. {\bf 273}, No. 4, 140 (1995), 
A. Ekert and R. Josza, Rev. Mod. Phys. {\bf 68}, 733 (1996).

\bibitem{bib_shor} P. W. Shor in {\it Proc. of the 35th Annual
Symposium on Foundations of Computer Science}, edited by S. Goldwasser
(IEEE Computer Society Press, New York, 1994), pp. 124-134.

\bibitem{bib_shannon} C. E. Shannon and W. Weaver, {\it The
mathematical theory of communication} (University of Illinois Press,
1949).

\bibitem{bib_schum} B. Schumacher, Phys. Rev. A {\bf 51}, 2738 (1995);
B. Schumacher and R. Josza, J. Mod. Optics {\bf 41}, 2343 (1994).
  
\bibitem{bib_shor1} P. W. Shor, Phys. Rev. A {\bf 52}, 2493 (1995); 

\bibitem{bib_calder} A. R. Calderbank and P.W. Shor, Phys. Rev. A {\bf
54}, 1098 (1996).

\bibitem{bib_steane} A. M. Steane, Proc. Roy. Soc. London (to be published);
Phys. Rev. Lett. {\bf 77}, 793 (1996);

\bibitem{bib_laflamme} R. Laflamme, C. Miquel, J. P. Paz, and
W. H. Zurek, Phys. Rev. Lett. {\bf 77}, 198 (1996)

\bibitem{bib_ekert} A. Ekert and C. Macchiavello,
Phys. Rev. Lett. {\bf 77}, 2585 (1996).
 
\bibitem{bib_bdsw} C. H. Bennett, D. DiVincenzo, J. A. Smolin, 
and W. K. Wootters, Phys. Rev. {\bf A 54}, 3824 (1996).

\bibitem{bib_knill} E. Knill and R. Laflamme, Phys. Rev. {\bf A 55},
  900 (1997).

\bibitem{bib_calder1} A. R. Calderbank, E. M. Rains, P. W. Shor, and 
N. J. A. Sloane, eprint quant-ph/9608006, to appear in IEEE Transactions
on Information Theory.

\bibitem{bib_schum1} B. Schumacher, Phys. Rev. {\bf A 54}, 2614 (1996).

\bibitem{bib_schum2} B. Schumacher and M.A. Nielsen, Phys. Rev. {\bf A
    54}, 2629 (1996).

\bibitem{bib_lloyd} S. Lloyd, Phys. Rev. {\bf A55}, 1613 (1997).

\bibitem{bib_hausladen} P. Hausladen, R. Josza, B. Schumacher,
M. Westmoreland, and W. K. Wootters, Phys. Rev. {\bf A 54}, 1869
(1996).

\bibitem{bib_kholevo97} A. S. Kholevo, eprint quant-ph/9611023.

\bibitem{bib_horo} R. Horodecki and M. Horodecki, Phys. Rev. {\bf A
    54}, 1838 (1996). 
    
\bibitem{bib_channel} C. H. Bennett, G. Brassard, S. Popescu,
B. Schumacher, J. A. Smolin, and W. K. Wootters, Phys. Rev. Lett. {\bf
76}, 722 (1996).


\bibitem{bib_ash} R. B. Ash, {\it Information Theory} (Dover, New York, 1965). 

\bibitem{bib_nocloning} W. K. Wootters and W. H. Zurek, Nature {\bf
299}, 802 (1982); D. Dieks, Phys. Lett. A {\bf 92}, 271 (1982).

\bibitem{bib_neginfo} N. J. Cerf and C. Adami, e-print
  quant-ph/9512022; N. J. Cerf and C. Adami, in Proc. of 2nd
  Intern. Symposium on Fundamental Problems in Quantum Physics, ed. by
  M. Ferrero and A. van der Merwe (Kluwer Academic Publishers,
  Dordrecht, 1997).
  
\bibitem{bib_entang} N. J. Cerf and C. Adami, Proc. 4th Workshop on
  Physics and Computation, Boston Univ. 22-24 Nov. 1996, T. Toffoli,
  M.  Biafore, and J. Leao, eds. (New England Complex Systems
  Institute, 1996), p. 65-71; e-print quant-ph/9605039.
 
\bibitem{bib_meas} N. J. Cerf and C. Adami, e-print quant-ph/9605002.

\bibitem{bib_reality} C. Adami and N. J. Cerf, Caltech preprint
KRL-MAP-204, August 1996.

\bibitem{bib_araki} H. Araki and E. H. Lieb, Comm. Math. Phys. {\bf
18}, 160 (1970). 

\bibitem{bib_wehrl} A. Wehrl, Rev. Mod. Phys. {\bf 50}, 221 (1978). 

\bibitem{bib_kholevo} A. S. Kholevo, Probl. Inform. Transmission {\bf
9}, 110 (1973).

\bibitem{bib_access} N. J. Cerf and C. Adami, eprint quant-ph/9611032.

\bibitem{bib_teleport} C. H. Bennett {\it et al.}, Phys. Rev. Lett. {\bf 70}, 
1895 (1993).

\bibitem{bib_smolin} P.W. Shor and J.A. Smolin, eprint quant-ph/9604006.

\bibitem{bib_superdense} C. H. Bennett and S. J. Wiesner,
  Phys. Rev. Lett. {\bf 69}, 2881 (1992).
\end{references}
\end{document}